\documentclass[english,conference]{IEEEtran}
\usepackage[latin9]{inputenc}
\usepackage{color}
\usepackage{array}
\usepackage{verbatim}
\usepackage{multirow}
\usepackage{amsmath}
\usepackage{amssymb}
\usepackage{graphicx}

\makeatletter

\providecommand{\tabularnewline}{\\}
\newcommand{\lyxdot}{.}

\usepackage{rotfloat}
\usepackage{cite}

\makeatother

\usepackage{babel}
\begin{document}
\title{Quantum circuit optimization for multiple QPUs using local structure}
\author{
\IEEEauthorblockN{Edwin Tham}
\IEEEauthorblockA{Entangled Networks Ltd.\\Toronto, ON, M4R 1A2, Canada\\
Email: edwin@entanglednetworks.com}\and
\IEEEauthorblockN{Ilia Khait}
\IEEEauthorblockA{Entangled Networks Ltd.\\Toronto, ON, M4R 1A2, Canada}\and
\IEEEauthorblockN{Aharon Brodutch}
\IEEEauthorblockA{Entangled Networks Ltd.\\Toronto, ON, M4R 1A2, Canada}
}
\maketitle
\begin{abstract}
Interconnecting clusters of qubits will be an essential element of scaling up future quantum computers. Operations between quantum processing units (QPUs) are usually significantly slower and costlier than those within a single QPU, so usage of the interconnect must be carefully managed. This is loosely analogous to the need to manage shared caches or memory in classical multi-CPU machines. Unlike classical clusters, however, quantum data is subject to the no-cloning theorem, which necessitates a rethinking of cache coherency strategies. Here, we consider simple strategies of using EPR-mediated remote gates and teleporting qubits between clusters as necessary -- generally expensive operations that we seek to minimize. Crucially, we develop optimizations at compile-time that leverage local structure in a quantum circuit, so as to minimize inter-cluster operations at runtime. We benchmark our approach against existing quantum compilation and optimization routines, and find significant improvements in circuit depth and interconnect usage.
\end{abstract}

\section{Introduction}

Quantum information processing (QIP) hardware have undergone rapid advances in both quality and quantity in recent years \cite{pelofske2022quantum,jurcevic2021,baldwin2022}. Nevertheless, most useful algorithms and routines that might be executed on a quantum computer continue to require resources that surpass the scale of current-generation quantum processing units (QPUs). The resource shortfall can be sheer number of qubits on a QPU (i.e. ``width''), number of quantum operations being performed before noise and errors accrued become overwhelming (i.e. ``depth''), or both. These resource shortfalls are potentially further exacerbated if error-correction schemes are employed.

Simply adding more qubits to a QPU is an obvious way to overcome width limitations. Yet in practice, we are prevented by technical obstacles from doing so indefinitely. In most QPU architectures, there are known limits beyond which, naively scaling up qubit-count yields diminishing returns. For instance, in trapped-ion QPUs adding more ions into a trap either increases the complexity or the duration of two-qubit gate operations, and necessitates cooling cycles that undo heating inadvertently introduced in the course of computation and read-out. In another example, solid-state QPUs (e.g. superconducting transmon qubits) grow in surface area with respect to the number of qubits it contains, inevitably running up against high substrate and manufacturing defect rates.

One avenue for increasing qubit count beyond the size limits inherent to many single-QPU systems is simply to interconnect multiple QPUs of bounded size. In classical computing, this is somewhat analogous to interconnecting ``chiplets'' -- usually classical processing units (CPUs) residing on monolithic silicon dies -- into a system-on-chip (SoC), or linking up fully functional computers with fast network interfaces. In either case, one ends up with a device with greater combined execution resources.

A crucial distinction when adopting a multi-QPU paradigm in quantum computing is that communication between QPUs must be mediated by quantum channels rather than classical ones. In some implementations, that quantum channel is realised simply by \emph{moving} a physical qubit from one QPU to another. However, quantum mechanics also allows for a pre-shared ``mediator'' or ``resource'' state (usually some highly entangled quantum state). This state is distributed between QPUs and, when it is used alongside \emph{classical} communication and teleportation-like mechanisms, can act as a quantum channel. Leveraging this fact, some multi-QPU schemes eschew physical transport of qubits between QPUs in favour of mechanisms for generating and distributing these resource states.

Properly considering and optimizing for the behaviour of these quantum channels requires careful compiler design. One aspect of the optimization is to ensure, as much as possible, that related data (e.g. those that must be operated on together) remain on the same QPU. This minimizes operations that straddle multiple QPUs, which are often slower and costlier than those that are resident on a single QPU.

\section{Background\label{sec:Background}}

\subsection{Flavours of quantum interconnect\label{subsec:InterconTypes}}

One method for interconnecting multiple QPUs is to transport a qubit physically. This process is generally slow and inefficient. We refer interested readers to \cite{pino2021-QCCD} for one example where trapped-ions are shuttled between traps in quantum charge-coupled devices (QCCD), and to \cite{tang2018-transducer,tang2021-transducer} for discussions on converting a microwave photon, which couples to a solid-state superconducting QPU, into an optical one that can be transported via fiber. With QCCDs, shuttling operations are $6-15$ times slower than operations ``native'' to a single QPU. Microwave-to-optical transducers on the other hand are lossy and currently are limited to single-digit efficiencies. For compilation purposes therefore, these transport operations represent a bottleneck to be minimized.

An alternative to physical transport is to leverage entangled resource states and classical communication\cite{kielpinski2002-ionIntercon}. Figure~\ref{fig:Interconnect} shows an example of two QPUs linked by an interconnect, whose function is simply to generate an entangled resource state. That resource state can be used to mediate interactions across the two QPUs using only \emph{local operations and classical communication} (LOCC). We will briefly describe how this works. Consider the so-called $\Phi^{+}$ Bell- or Einstein-Podolsky-Rosen (EPR) state:
\begin{figure}[t]
\begin{centering}
\includegraphics[width=0.48\textwidth]{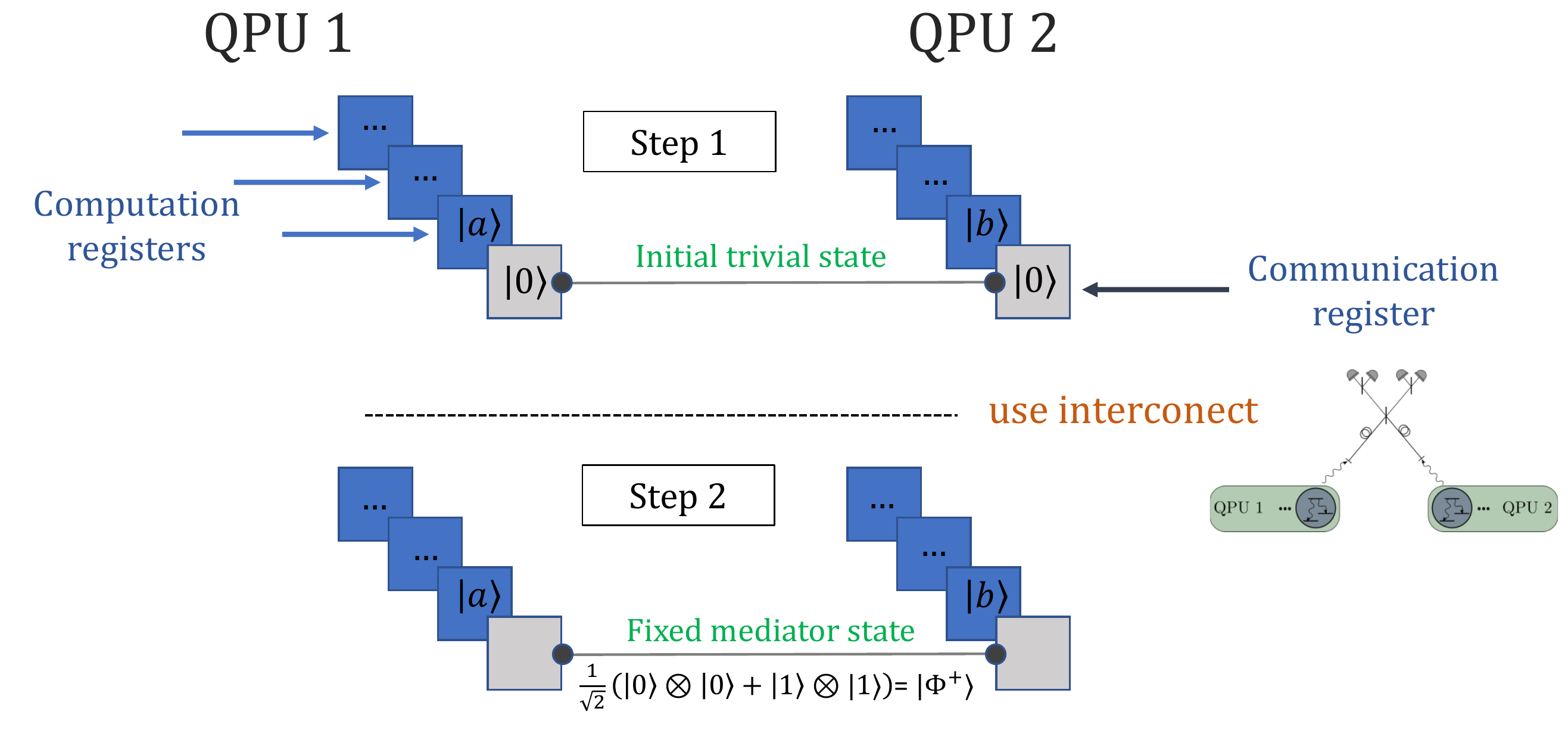}
\par\end{centering}
\caption{Illustration of an interconnect being used to generate a resource state between two separate quantum processing units (QPUs). In this example each QPU contains a number of data qubits (blue) and one interconnect qubit (gray). The interconnect (right hand figure) collects a single photon from each interconnect qubit and interferes it in order to entangle the two interconnect qubits. The resulting entangled state can be used as a resource to enable quantum communication.\label{fig:Interconnect}}
\end{figure}
\begin{equation}
\left|\Phi^{+}\right>=\frac{1}{\sqrt{2}}\sum_{k\in\{0,1\}}\left|k\right>_{1}\left|k\right>_{2}\label{eq:nolabel1}
\end{equation}
The susbscript in each ket in the summand indicates that the ket addresses states in QPU 1 (or 2) respectively. Now suppose one is interested in performing a coherent XOR (also known as a controlled-$\sigma_{X}$ or ``CNOT'') operation, defined as:
\begin{equation}
\text{XOR}_{a,b}:\left|a\right>\left|b\right>\mapsto\left|a\right>\left|a\oplus b\right>\label{eq:CNOTideal}
\end{equation}
where $\oplus$ denotes addition modulo 2. Supposing the states $\left|a\right>$ and $\left|b\right>$ reside on QPUs 1 and 2 respectively, the desired XOR operation stipulated here can be mediated by that $\left|\Phi^{+}\right>$ resource state as follows. Start with the initial state: 
\begin{align}
\left|\psi\right> & =\left|a\right>_{1}\left|\Phi^{+}\right>\left|b\right>_{2}\nonumber \\
 & =\left|a\right>_{1}\otimes\left[\frac{1}{\sqrt{2}}\sum_{k\in\{0,1\}}\left|k\right>_{1}\left|k\right>_{2}\right]\otimes\left|b\right>_{2}\label{eq:nolabel2}
\end{align}
Do an XOR between $\left|a\right>_{1}$ and 2nd register (half of $\Phi^{+}$ that resides on QPU 1); as well as between the 3rd register (half of $\Phi^{+}$ that resides on QPU 2) and $\left|b\right>_{2}$:
\begin{align}
\text{XOR} & :\left|\psi\right>\nonumber \\
 & \mapsto\left|a\right>_{1}\otimes\left[\frac{1}{\sqrt{2}}\sum_{k\in\{0,1\}}\left|k\oplus a\right>_{1}\left|k\right>_{2}\right]\otimes\left|b\oplus k\right>_{2}\label{eq:nolabel3}
\end{align}
Now suppose we measure the 2nd register and find the \emph{classical} bit $k\oplus a=m$, which implies $k=a\oplus m$. The rules of quantum measurements implies that the quantum state immediately collapses to:
\begin{align}
 & \left|a\right>_{1}\otimes\left[\left|k\oplus a\right>_{1}\left|k\right>_{2}\right]\otimes\left|b\oplus k\right>_{2}\nonumber \\
= & \left|a\right>_{1}\otimes\left[\left|m\right>_{1}\left|a\oplus m\right>_{2}\right]\otimes\left|b\oplus a\oplus m\right>_{2}\label{eq:nolabel4}
\end{align}
If QPU 1 can \emph{classically} communicate the bit $m$ with QPU 2, then the latter can perform a bit-flip conditioned upon that bit, which yields:
\begin{align}
 & \left|a\right>_{1}\otimes\left[\left|m\right>_{1}\left|a\oplus m\right>_{2}\right]\otimes\sigma_{X}^{m}\left|b\oplus a\oplus m\right>_{2}\label{eq:nolabel5a}\\
= & \left|a\right>_{1}\otimes\left[\left|m\right>_{1}\left|a\oplus m\right>_{2}\right]\otimes\left|b\oplus a\right>_{2}\label{eq:nolabel5b}
\end{align}
Finally, we can apply a Hadamard operator to the 3rd register and measure it (obtaining a classical bit $n$) to disentangle remnants of the EPR-state, leaving us with:
\begin{equation}
(-1)^{(a\oplus m)n}\left|a\right>_{1}\otimes\left|b\oplus a\right>_{2}\label{eq:nolabel6}
\end{equation}
The stray phase here can be corrected by applying $\sigma_{Z}^{n}$, which QPU 1 can do if the bit $n$ is communicated by QPU 2. Note, the result is the desired output of the original XOR operation defined in \ref{eq:CNOTideal}. We had effected this XOR between qubits on separate QPUs by consuming an EPR state and exchanging only one pair of classical bits between QPUs 1 and 2.

A host of other constructions that use EPR resource states to yield common operations (e.g. bit-swap, Tofolli, etc.) can similarly be derived. Particularly noteworthy is basic teleportation \cite{bouwmeester1997-teleportation}, which most closely mimics simply moving a qubit like physical transport would.

Ultimately, key to being able to perform any EPR-mediated operations between QPUs is the ability to produce and distribute the resource states to begin with. To that end, optical schemes had been demonstrated that can generate EPR-states between trapped ions at up to $\sim200$~Hz\cite{monroe2014-interconnects,monroe2013-reviewinterconnects,awschalom2021-reviewQUICS}. This is faster than some QCCD shuttling operations, but is still very slow compared to native single-QPU operations\cite{linke2017-QPUcomparison}.

\subsection{Quantum compilation\label{subsec:QuantumCompilation}}

The essential remit of a compiler is to reconstruct a program to be better suited for execution on hardware, while preserving its functionality. In the context of quantum computing, programs are often specified as quantum circuits constructed from basic gates. The quantum compiler must ensure that all gates specified within the quantum circuit are supported in hardware. For single-qubit gates, this is a matter of finding decompositions of a unitary matrix in $U(2)$ in a hardware-compatible basis.

Gates involving more qubits (like the XOR discussed in Section~\ref{subsec:InterconTypes}) incur an additional complication: Many QPUs do \emph{not} support two-qubit operations between arbitrary qubit pairs. This is usually conveyed succinctly as a ``coupling map'', an undirected graph $G_{\text{cmap}}$ where vertices represent qubits and edges represent supported two-qubit operations. If a desired gate (say a $U(4)$ operation) addresses a pair of qubits that are not directly coupled in hardware (i.e. vertices in $G_{\text{cmap}}$ not connected by an edge), a simple basis change is insufficient. Instead, a common tack is to find an indirect route between relevant vertices in $G_{\text{cmap}}$, and then re-synthesize the desired gate in terms of supported operations along that route. If $G_{\text{cmap}}$ is disjoint, that is some vertices are simply unreachable from some others, then certain two-qubit (or multi-qubit) operations are simply impossible on that hardware.

This process of resolving two-qubit gates onto a given hardware topology can result in many additional operations being introduced (${\cal O}(4d)$ for remote operations, ${\cal O}(3d)$ for a swap-based approach \emph{without} reverse swap; here $d$ is length of the indirect route between qubits targeted by the two-qubit operation). The resulting hardware-mapped circuit therefore can have significantly increased depth\cite{zulehner2018-IBMcnotremap}. Much work had been devoted to finding hardware-efficient two-qubit gate re-synthesis, given a circuit\cite{amy2018-cnotsynth,kissinger2019-cnotsynth,nash2020-cnotsynth}. Slightly less obvious is the fact that the number of additional gates introduced during re-synthesis can vary wildly depending on how qubits in a quantum circuit (i.e. ``logical'' qubits) are assigned to particular vertices in $G_{\text{cmap}}$ (i.e. ``physical'' qubits). There is every chance that a naive assignment (e.g. simply assigning logical qubit $0\to$ physical qubit $0$, $1\to1$, $2\to2$, and so on) will be sub-optimal. An important aspect of quantum compilation therefore includes the construction of a logical to physical qubit map that yields an efficient circuit.

Unfortunately, finding an optimal assignment is potentially a hard classical problem. The space of possible logical $\to$ physical qubit maps is $N!/(N-M)!$, where $N$ ($M$) is the total number of physical (logical) qubits, which rules out a brute-force search for large circuits or QPUs. More formally, suppose we construct a coupling map analogous to $G_{\text{cmap}}$ but specified \emph{not} by hardware topology but rather from two-qubit gate density in a quantum circuit; call this graph $G_{\text{circ}}$. Then, finding the optimal assignment in some cases reduces to finding a graph isomorphism between $G_{\text{circ}}$ and subgraphs of $G_{\text{cmap}}$. This is generally NP-complete and not tractable when $N$ and $M$ are large and neither graph is very sparse or disjoint\cite{carletti2017-VF3complexity,carletti2017-VF3}. In the rest of the manuscript, we will describe ``gentler'' graph problems relevant to our compilation approach.

\section{Compilation Technique}

\subsection{Choice of Topology\label{subsec:ChoiceofTopology}}

While the qubit assignment problem (mapping logical $\to$ physical qubits) generally construed in Section~\ref{subsec:QuantumCompilation} is not tractable, particular instances of the problem \emph{may} be satisfactorily solved in practice\cite{carletti2017-VF3,carletti2017-VF3complexity}. For our purposes in this manuscript however, we will avoid the isomorphism problem entirely by focusing only on hardware topologies that are ``well clustered''. These are characterised by graphs that admit clusters of qubits that possess denser intra-cluster couplings and comparatively fewer inter-cluster ones.

Given the context of multi-QPU architectures that we are considering, this sort of hardware topology is natural. In Section~\ref{sec:Background} we described various realisations of inter-QPU operations, all of which are bottlenecked in rate and cost given currently available hardware. The assumption, that interconnects will remain a scarce resource within multi-QPU architectures, is likely a reasonable one for the foreseeable future. An extreme example is embodied in monolithic ion-trap devices in which native intra-QPU couplings are fundamentally all-to-all (so that $G_{\text{cmap}}$ generally has $\mathcal{O}(N^{2})$ edges), while inter-QPU couplings either through distributed EPR-pairs or shuttled ions will likely remain sparse (possibly $\mathcal{O}(N)$) due to rate limits and cost of multiplexing many simultaneous Hong-Ou-Mandel interactions\cite{monroe2013-reviewinterconnects}. In limiting our focus to ``well-clustered'' hardware topologies, the central thesis is that it becomes far more important optimize for the expensive or scarce resource (i.e. usage of inter-QPU couplings), rather than any and all couplings more generally.

To be more precise, how ``well-clustered'' a given coupling map is can be evaluated using several measures, one being the graph \emph{conductance} $\phi_{G}$. Supposing one attempts to partition the graph $G_{\text{cmap}}$ into $k$ disjoint clusters corresponding to separate QPUs, then for $1\leq j\leq k$, conductance is defined as:
\begin{equation}
\phi_{G}(j)=\min_{\vec{v}}\frac{\vec{v}_{j}^{T}(D-A)\vec{v}_{j}}{\vec{v}_{j}^{T}D\vec{v}_{j}},\label{eq:nolabel7}
\end{equation}
for our purposes, $A$ is the adjacency matrix which $i,j$-th entry is $1$ if the $i$-th and $j$-th vertices are joined by an edge; $D$ is a degree matrix s.t. $D_{ij}=\delta_{ij}\sum_{k}A_{jk}$; and $\vec{v}_{j}$ is an indicator vector whose $k$-th entry is $1$ if the $k$-th vertex belong in the $j$-th cluster and is $0$ otherwise. A well-clustered graph admits a partitioning s.t. $\phi_{G}(j)$ is low for all $j$. As a means for quickly evaluating whether a hardware topology is well-suited for our compilation approach, $\phi_{G}$ is convenient since it is known to be upper-bounded by the $k$-th eigenvalue of the Laplacian $\mathcal{L}=D-A$ through an extension to Cheeger's inequality\cite{trevisan2014-higherordercheeger}. The eigenvalues of $\mathcal{L}$, in turn, can be computed directly without first having to \emph{find} the partitions that specify $\vec{v}$. Another important measure for our purposes is the association ratio:
\begin{equation}
R_{G}(j)=\frac{\vec{v}_{j}^{T}A\vec{v}_{j}}{\vec{v}^{T}\vec{v}}\label{eq:nolabel8}
\end{equation}

We consider our methods to be applicable for hardware topologies that exhibit small $\phi_{G}$, and \emph{especially applicable} where $\phi_{G}$ decreases with increasing qubit count $N$ and roughly static $R_{G}$.

\subsection{Global QPU assignment\label{subsec:Global}}

One consequence of choosing to optimize inter-QPU operations, is that we are now concerned mainly with the problem of assigning logical qubits to \emph{QPU}s rather than to physical qubits. Suppose, from an input quantum circuit, we construct a coupling graph ($G_{\text{circ}}$) s.t. its adjacency matrix entry $A_{ij}$ is simply the number of 2-qubit gates between qubits $i$ and $j$. Without loss of generality assume that there are no multi-qubit gates beyond two-qubit ones\footnote{If an input circuit were to contain multi-qubit gates, universality results ensure that we can always efficiently decompose them into more elementary blocks consisting of single- and two-qubit gates at most\cite{dawson2005-SKDN,gottesman1998-heisenberg}}. Then, minimizing the number of inter-QPU operations in a $k$-QPU architecture is equivalent to minimizing the cost function:
\begin{equation}
C_{\text{KL}}=\sum_{j=1}^{k}\vec{v}_{j}^{T}{\cal L}_{\text{circ}}\vec{v}_{j},\label{eq:nolabel9}
\end{equation}
where ${\cal L}$ is the Laplacian matrix for $G_{\text{circ}}$ as defined in Section~\ref{subsec:ChoiceofTopology}, and the indicator vector $\vec{v}_{j}$ takes value $1$ in its $r$-th entry if the $r$-th logical qubit is assigned to QPU-$j$, and $0$ otherwise. Finding $\vec{v}_{j}$'s that minimize $C_{\text{KL}}$ is precisely equivalent to solving the so-called ``minimum-cut'' problem.

Since the size of QPUs are usually fixed, an additional constraint must be added during minimization of $C_{\text{KL}}$:
\begin{equation}
\vec{v}_{j}^{T}\vec{v}_{j}=\text{Size of QPU }j.\label{eq:nolabel10}
\end{equation}
Generally, a cardinality-constrained ``minimum-cut'' problem is substantially harder than the unconstrained variant. However, approximate solvers can efficiently generate satisfactory solutions in practice\cite{karypis1997-metis,schlag2016-kahypar,dhillon2007-graphcuts}. A technique that we adopt and implement here is spectral partitioning. Consider the case where the sizes of all QPUs are the same, i.e. $\vec{v}_{j}^{T}\vec{v}_{j}=s$ is constant for all $j$, and the corresponding minimization problem is of the ``Kernighan-Lin'' variety\cite{kernighan1970}. Since $\mathcal{L}=D-A$ is Hermitian, the Courant-Fisher theorem implies that orthonormal eigenvectors ($\vec{u}_{j}=\vec{v}_{j}/\sqrt{s}$) of $s\mathcal{L}$ corresponding to the first $k$ lowest eigenvalues minimizes $C_{\text{KL}}$ while satisfying the cardinality constraint\cite{ikebe1987-minmaxtheorems}. While the theorem holds strictly only when $\vec{u}_{j}\in\mathbb{C}^{N}$, in practice spectral partitioning nevertheless yields good approximate solutions when entries in $\sqrt{s}\vec{u}_{j}$ must be rounded to $0$ or $1$. Partitioning in the case of clusters of unequal sizes have also been studied\cite{dhillon2007-graphcuts,satuluri2009-MLRMCR}.

Once a good partitioning is found, we assign logical qubits from the input quantum circuit to a random qubit within the target QPU (as specified by $\{\vec{v}_{j}\}$), whereupon a more traditional swap insertion method as described in Section~\ref{subsec:QuantumCompilation} is used if any one QPU has an internal topology sparser than all-to-all. Note, while we use the language of inter- and intra-QPU here, the preceding discussion is applicable even if the architecture does not strictly speaking contain interconnected QPUs, so long as it satisfies the conditions laid out in Section~\ref{subsec:ChoiceofTopology}.

For the interested reader, previous (static/global) qubit mapping strategies have been proposed and implemented~\cite{xieSABRE2021,harrow2021}. Our approach differs in that it focuses on the minimization of inter-QPU operations specifically, so we are able to leverage efficient graph-CUT solvers to better handle large QPUs/circuits.

\subsection{Local optimization\label{subsec:Local}}

In Section~\ref{subsec:Global}, the coupling map $G_{\text{circ}}$ had been constructed by considering \emph{all} two-qubit gates in the quantum circuit. For an \emph{initial} assignment of logical qubits to QPUs, it is reasonable to look at the circuit as a whole. However, such a construction strips out the chronological order in which operations are performed during execution of the quantum circuit. Since most quantum gates are \emph{not} commutative (so chronological order matters), and because a significant source of bottlenecks is slow inter-QPU operation times (see Section~\ref{subsec:InterconTypes}), it is important to ensure that heuristic we might use for that optimization takes into account temporal structures and correlations in the program being executed.

To give a simple concrete example, suppose a quantum circuit prescribes many two-qubit gates between qubits $q_{1}$ and $q_{2}$ as well as between $q_{1}$ and $q_{3}$. Globally, this implies a strong preference for $q_{1}$, $q_{2}$, and $q_{3}$ all to be clustered. However, if $q_{1}$ and $q_{3}$ interactions are chronologically localized to times \emph{much later} than $q_{1}$ and $q_{2}$ interactions (suppose they are temporally separated by much more than relevant hardware timescales), then there is no longer any reason to insist that $q_{2}$ and $q_{3}$ be clustered -- there is no cost to $q_{2}$ being ``far away'' by the time $q_{1},q_{3}$ couplings need to be realised.

In order to exclude spurious clustering constraints from interactions that are temporally too far separated, we elected to re-construct $G_{\text{circ}}$ based only on a subset of operations in the quantum circuit. To do this, we reasonably assume that the target hardware is sufficiently well-characterized such that operation times for various gates are known. That way, each operation stipulated in the quantum circuit can be assigned an \emph{expected} time-to-execution, $t_{\text{gate}}$. We then specify a rolling time interval, $[t_{\text{start}},t_{\text{start}}+\Delta t)$. We then construct a $G_{\text{circ}}$ by considering \emph{only} gates that occur within that rolling window (i.e. $t_{\text{start}}\leq t_{\text{gate}}\leq t_{\text{start}}+\Delta t$). That circuit graph is then subjected to the same minimum cut treatment as in Section~\ref{subsec:Global}.

Discussions of specific techniques for selecting optimal parameters that define the rolling window is outside the scope of this manuscript. For the purposes of ensuing experimental results, one effective choice is simply to select $\Delta t$ to correspond to the largest limiting timescale of the target hardware, such that the availability of execution resources (like distributed EPR-states) is unlikely to depend on operations performed more than $\Delta t$ ago. In turn, a set $t_{\text{start}}$ can simply be chosen so as to produce disjoint time windows that cover the quantum circuit from start to end of execution.

Once an approximate minimum-cut is found on a (temporally) localized $G_{\text{circ}}$, we can compare the new QPU assignment it implies to the existing one (for the first rolling window, the existing assignment is the one found from the global graph in Section~\ref{subsec:Global}). A decision is then made as to whether to effect a QPU \textbf{\emph{re-}}assignment (by means of EPR-mediated teleportation, swap or physical transport) or to simply leave the current assignment as-is relying instead on logical swap insertions and/or EPR-mediated CNOT (see Section~\ref{subsec:InterconTypes}).

We point out that for small $\Delta t$ and/or very deep quantum circuits, one may have to perform partitioning of many graphs. Fortunately, graphs corresponding to distinct time windows can be partitioned in embarrassingly parallel fashion. Since most quantum algorithms are useful only when they have bounded runtime, the number of time windows in our local optimization scheme is similarly bounded.

\section{Benchmark}

\subsection{Methodology}

In order to test our optimization methods, we implemented the techniques discussed in Section~\ref{subsec:Local} as a Python/NumPy/Numba library (referred to as ``MultiQopt'' below). In order to standardize inputs, we implemented interfaces to the IBM's QISkit software development kit (SDK) so that quantum circuits can be defined in terms of QISkit\cite{qiskit} circuit objects or as Quantum Assembly (QASM) 2.0 strings\cite{cross2017-openQASM}. Since graph partitioning is central to our technique, our implementation can call external partitioners like KaHyPar\cite{schlag2016-kahypar} and MeTiS\cite{karypis1997-metis} as well as an internal spectral partitioner.

In addition to a quantum circuit, the target hardware topology ($G_{\text{cmap}}$) can be optionally supplemented with a ``role-assignment'' for various physical qubits. Among other things, physical qubits can be explicitly assigned to particular QPU objects, and qubits can can be earmarked for ``special use'' like holding an EPR state. The output of our implementation is an optimized QISkit circuit object, along with logical-to-physical qubit assignment lookup tables and possibly custom QISkit instructions for EPR-mediated operations.

We compiled quantum routines from a suite of benchmark circuits\cite{feynman,meamy2019-Feynman}, targeting an architecture consisting of two clusters of all-to-all coupled QPUs interconnected with two EPR-mediated interconnects. Given an input quantum circuit occupying $N$-qubits, we set the size of each target QPU to $\lceil N/2\rceil+2$ (the added two qubits on each QPU serves as an EPR-state reservoir). This architecture choice is deliberate as it falls in the regime where assumptions behind our optimizations hold. But more importantly, it is representative of an architecture that trapped-ion quantum computer builders foresee in the near-future\cite{awschalom2021-reviewQUICS}.

For comparison, we also compiled the same circuits targeting the same architecture with QISkit (using compile option ``optimization\_level=3'', stipulating maximum optimizations at the expense of longer compile times)\cite{qiskit}. In all cases, the target ``native'' gates were selected to be the set $\{r_{x},r_{z},h,cx\}$. Here, $cx$ is the CNOT operation defined in \ref{eq:CNOTideal} and:
\begin{align}
r_{x}(2\theta) & =e^{i\sigma_{x}\theta}=\left[\begin{array}{cc}
\cos\theta & i\sin\theta\\
i\sin\theta & \cos\theta
\end{array}\right]\label{eq:nolabel11}\\
r_{z}(2\theta) & =e^{i\sigma_{z}\theta}=\left[\begin{array}{cc}
e^{-i\theta} & 0\\
0 & e^{i\theta}
\end{array}\right]\label{eq:nolabel12}\\
h & =\frac{1}{\sqrt{2}}\left[\begin{array}{cc}
1 & 1\\
1 & -1
\end{array}\right].\label{eq:nolabel13}
\end{align}

The benchmark environment is an AMD Ryzen 5 5600X machine with 16GB DDR4-3600 memory, with Arch Linux kernel 5.15.1. Some additional development environment information include Python-3.8, NumPy-1.21.1, Numba-0.54, and QISkit-0.26. The outputs of all compilers were then analyzed for (a) total number of two-qubit gates (b) total number of ``expensive'' interconnect uses, here defined to be any two-qubit operation that spanned the two all-to-all coupled qubit clusters and (c) compilation time.

\subsection{Results}

\begin{figure}[t]
\begin{centering}
\includegraphics[width=0.45\textwidth]{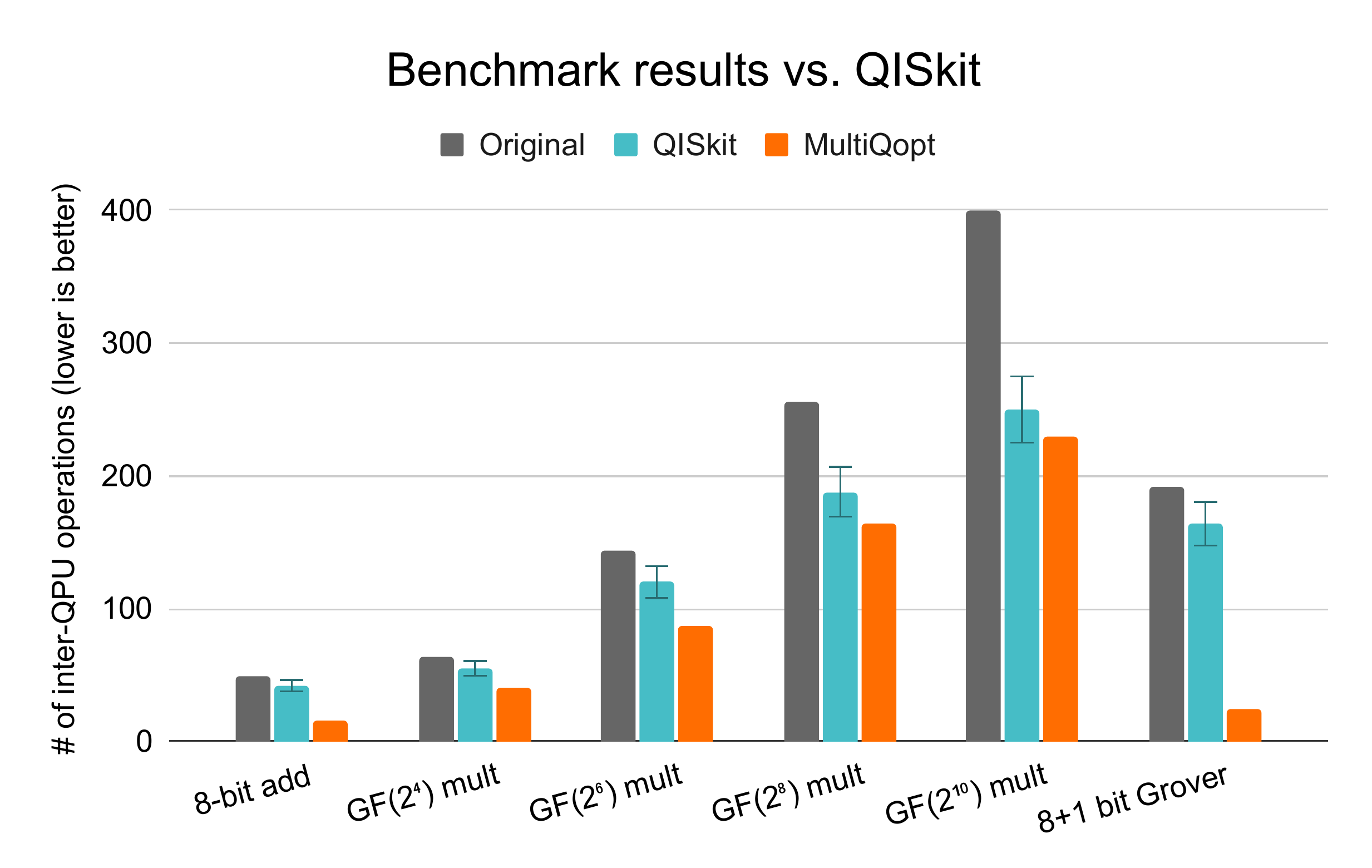}
\par\end{centering}
\caption{Benchmark result showing number of inter-cluster operations, comparing MultiQopt to QISkit (ver. 0.26). The original circuit (gray) which was compiled onto an all-to-all connected architecture is re-compiled onto the target dual QPU architecture using QISkit (turqoise) and MultiQopt (orange). Compilation with MultiQopt shows a significant reduction in the number of inter-cluster operations.\label{fig:ResultFig}}
\end{figure}
\begin{figure}[t]
\begin{centering}
\includegraphics[width=0.45\textwidth]{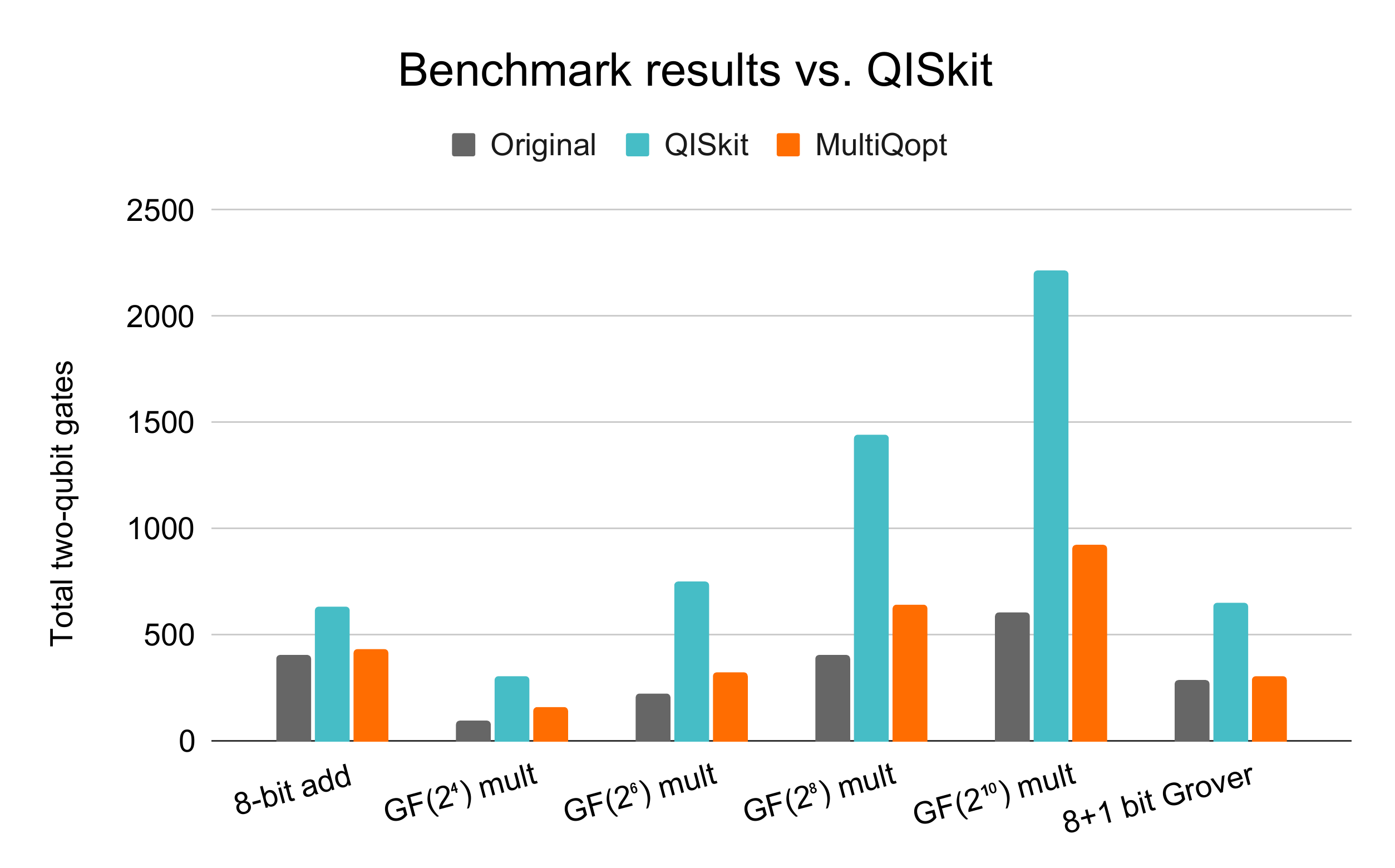}
\par\end{centering}
\caption{Benchmark result showing total two-qubit gates, comparing MultiQopt to QISkit (ver. 0.26). The original circuit (gray) which was compiled onto an all-to-all connected architecture is re-compiled onto the target dual QPU architecture using QISkit (turqoise) and MultiQopt (orange). Compilation with MultiQopt shows a significant reduction in the number of two qubit operations.\label{fig:ResultFig-tot}}
\end{figure}

We show results for several familiar circuits (an adder, multipliers of various sizes, a Grover routine) in Figure~\ref{fig:ResultFig} and Tables~\ref{tab:BenchResults-1a}~-~\ref{tab:BenchResults-1b}. Benchmark results for additional input programs are shown in the Appendix. Additionally, current and more extensive sets of benchmark results can be found online at \cite{multiqopt}.

In Figure~\ref{fig:ResultFig}, we see a clear improvement over QISkit in the quantity of interest, the total number of interconnect uses. Additionally, we also show in Figure~\ref{fig:ResultFig-tot} that this gain is \emph{not} won at the cost of total two-qubit gates -- the latter remains significantly lower with our approach compared to QISkit. We attribute this behaviour to the fact that our approach is naturally suited to mapping classes of graphs that are ``clustered'' as we've described in Section~\ref{subsec:ChoiceofTopology}, whereas QISkit's heuristic algorithm, SABRE, isn't. Table~\ref{tab:BenchResults-1a} lists additional details about each input program in their \emph{uncompiled} form as read from their respective QASM files. Logical qubits are assumed to map to physical ones in \emph{naive} fashion; since no attempt is made to map to the target hardware topology, total two-qubit gates tends to be lower in the ``original'' input circuit owing to the lack of additional swap operations being inserted (see Section~\ref{subsec:QuantumCompilation}). Table~\ref{tab:BenchResults-1a-EN} shows results for our MultiQopt optimizer. The column labelled ``Global pass'' shows results with just the initial global QPU assignment pass (Section~\ref{subsec:Global}) whereas the column labelled ``local optimization'' shows the full compilation run that includes local optimizations (Section~\ref{subsec:Local}). The former does not report a separate column of total two-qubit gates because by construction, that is \emph{not }affected by the global pass. Finally, Table~\ref{tab:BenchResults-1b} shows corresponding results for QISkit (ver. 0.26). Since QISkit outputs exhibit significant run-to-run variation in inter-QPU two-qubit operations, we also report the spread, aggregating results over 30 repeated runs.
\begin{table}[t]
\centering{}\caption{\label{tab:BenchResults-1a}Benchmark baselines, showing two-qubit gate counts on several subroutines as defined in their respective input QASM files. {*}The baseline ``inter-QPU'' count is based on a trivial map between logical and physical qubits.}
\begin{tabular}{|c||c|c|c|}
\hline 
 &  & \multicolumn{2}{c|}{Base}\tabularnewline
\hline 
 & \#-qb & Total 2-qb & InterQPU{*}\tabularnewline
\hline 
\hline 
8-bit add & 24 & 409 & 49\tabularnewline
\hline 
GF($2^{4}$) mult & 12 & 99 & 64\tabularnewline
\hline 
GF($2^{6}$) mult & 18 & 221 & 144\tabularnewline
\hline 
GF($2^{8}$) mult & 24 & 405 & 256\tabularnewline
\hline 
GF($2^{10}$) mult & 30 & 609 & 400\tabularnewline
\hline 
8+1 bit Grover & 9 & 288 & 192\tabularnewline
\hline 
\end{tabular}
\end{table}
\begin{table}[t]
\centering{}\caption{\label{tab:BenchResults-1a-EN}Benchmark results for MultiQopt}
\begin{tabular}{|c||c|c|c|c|}
\hline 
\multirow{2}{*}{\textcolor{black}{MultiQopt}} & \multicolumn{1}{c|}{\textcolor{red}{Global pass}} & \multicolumn{2}{c|}{\textcolor{red}{w/ local optimization}} & \multirow{2}{*}{\textcolor{red}{Runtime}}\tabularnewline
\cline{2-4} \cline{3-4} \cline{4-4} 
 & \textcolor{red}{InterQPU} & \textcolor{red}{Total 2-qb} & \textcolor{red}{InterQPU} & \tabularnewline
\hline 
\hline 
8-bit add & \textcolor{red}{22} & \textcolor{red}{433} & \textcolor{red}{16} & \textcolor{red}{0.73s}\tabularnewline
\hline 
GF($2^{4}$) mult & \textcolor{red}{49} & \textcolor{red}{157} & \textcolor{red}{41} & \textcolor{red}{0.41s}\tabularnewline
\hline 
GF($2^{6}$) mult & \textcolor{red}{109} & \textcolor{red}{329} & \textcolor{red}{87} & \textcolor{red}{1.47s}\tabularnewline
\hline 
GF($2^{8}$) mult & \textcolor{red}{199} & \textcolor{red}{646} & \textcolor{red}{164} & \textcolor{red}{4.36s}\tabularnewline
\hline 
GF($2^{10}$) mult & \textcolor{red}{301} & \textcolor{red}{921} & \textcolor{red}{229} & \textcolor{red}{8.82s}\tabularnewline
\hline 
8+1 bit Grover & \textcolor{red}{48} & \textcolor{red}{304} & \textcolor{red}{24} & \textcolor{red}{1.13s}\tabularnewline
\hline 
\end{tabular}
\end{table}
\begin{table}[t]
\centering{}\caption{\label{tab:BenchResults-1b}Benchmark results for QISkit 0.26 (aggregated over 30 runs).}
\begin{tabular}{|c||c|c|c|}
\hline 
 & \multicolumn{3}{c|}{\textcolor{blue}{Qiskit 0.26 transpile (opt=3)}}\tabularnewline
\hline 
 & \textcolor{blue}{Total 2-qb} & \textcolor{blue}{InterQPU} & \textcolor{blue}{Runtime}\tabularnewline
\hline 
\hline 
8-bit add & \textcolor{blue}{630} & \textcolor{blue}{$42\pm12$} & \textcolor{blue}{2.69s}\tabularnewline
\hline 
GF($2^{4}$) mult & \textcolor{blue}{310} & \textcolor{blue}{$55\pm8$} & \textcolor{blue}{1.34s}\tabularnewline
\hline 
GF($2^{6}$) mult & \textcolor{blue}{757} & \textcolor{blue}{$120\pm15$} & \textcolor{blue}{2.75s}\tabularnewline
\hline 
GF($2^{8}$) mult & \textcolor{blue}{1444} & \textcolor{blue}{$188\pm14$} & \textcolor{blue}{5.12s}\tabularnewline
\hline 
GF($2^{10}$) mult & \textcolor{blue}{2217} & \textcolor{blue}{$250\pm25$} & \textcolor{blue}{7.83s}\tabularnewline
\hline 
8+1 bit Grover & \textcolor{blue}{652} & \textcolor{blue}{$164\pm27$} & \textcolor{blue}{2.77s}\tabularnewline
\hline 
\end{tabular}
\end{table}

\begin{comment}
IEEE bib:

\textbackslash bibliographystyle\{IEEEtran\} \textbackslash bibliography\{IEEEabrv,bib\}
\end{comment}

\section{Discussion and Conclusion}

We have discussed a multi-QPU centric quantum circuit compilation and optimization approach, premised upon the idea that inter-QPU operations are likely to remain expensive and scarce for the foreseeable future, despite being essential to the serious scaling up of quantum computers. Recognizing local structures in input quantum circuits allows for more flexible optimizations. Our approach also reduces to well-understood graph theoretic problems that admit approximate solutions that can be found efficiently for large classes of common graphs.

When targeting architectures likely to be typical in multi-QPU architectures in the near future, our benchmarks indicate our optimization approach yields significantly simpler circuits. Despite its multi-QPU centric background however, well-known bounds suggest that even monolithic QPU architectures with fairly common topologies may well benefit from our methods; but we leave benchmarking of these alternate topologies for a future work.

Some readers may recognize optimizations like mid-execution QPU re-assignment as being vaguely analogous to runtime optimizations that aim to maximize cache coherency and residency in multi-CPU classical architectures. Indeed, they share the common goal of minimizing QPU (or CPU) idle time by attempting to ensure relevant data is nearby when or where they at needed. Unlike classical measures, however, quantum information possess uniquely quantum idiosyncrasies; the no-cloning theorem for example implies that except in very narrow circumstances quantum data almost always needs to be \emph{moved}, \emph{not copied}. But QIP also allows for the use of shared resource states that can be used after-the-fact to generate large arbitrary entangled states without the various QPUs having to interact any further, but for the exchange of \emph{classical} bits.

All of this necessitates careful (re-)thinking of optimization strategies for multi-QPU systems. The present manuscript represents a promising step in that direction.

\bibliographystyle{IEEEtran}
\bibliography{IEEEabrv,bib}

\section*{Appendix A: Additional benchmark results.}

Tab.~\ref{tab:BenchResultsExtra-1} shows benchmark results for a large variety of other input quantum programs and sub-routines. As described in the main text, we show MultiQopt compared with QISkit 0.26, called with the same target hardware topologies.

\begin{sidewaystable*}
\begin{centering}
\begin{tabular}{|c||c|c|c|c|c|c|c|c|c|c|c|}
\hline 
 &  & \multicolumn{2}{c|}{Base} & \multicolumn{3}{c|}{\textcolor{blue}{Qiskit 0.26 transpile (opt=3)}} & \multicolumn{2}{c|}{\textcolor{red}{EN, global pass only}} & \multicolumn{3}{c|}{\textcolor{red}{EN, with local optimization}}\tabularnewline
\hline 
 & \#-qb & Total 2-qb & InterQPU & \textcolor{blue}{Total 2-qb} & \textcolor{blue}{InterQPU} & \textcolor{blue}{Runtime} & \textcolor{red}{InterQPU} & \textcolor{red}{Runtime} & \textcolor{red}{Total 2-qb} & \textcolor{red}{InterQPU} & \textcolor{red}{Runtime}\tabularnewline
\hline 
\hline 
Barenco (x3) & 5 & 24 & 16 & 41 & $13\pm4$ & 0.289s & 8 & 0.006179s & 28 & 4 & 0.061137s\tabularnewline
\hline 
Barenco (x4) & 7 & 48 & 32 & 94 & $25\pm16$ & 0.631s & 16 & 0.011019s & 56 & 8 & 0.163377s\tabularnewline
\hline 
Barenco (x5) & 9 & 72 & 48 & 156 & $44\pm11$ & 1.060936s & 16 & 0.016897s & 76 & 8 & 0.260133s\tabularnewline
\hline 
Barenco (x10) & 19 & 192 & 128 & 465 & $130\pm25$ & 3.395s & 16 & 0.060116s & 196 & 12 & 0.75719s\tabularnewline
\hline 
QCLA (mod 7) & 26 & 382 & 69 & 794 & $77\pm34$ & 5.676s & 56 & 0.069019s & 392 & 32 & 2.021352s\tabularnewline
\hline 
QCLA (com 7) & 24 & 155 & 24 & 212 & $10\pm6$ & 1.962s & 12 & 0.026569s & 188 & 10 & 0.345878s\tabularnewline
\hline 
QSLA (mux 3) & 15 & 80 & 26 & 155 & $22\pm4$ & 1.356s & 15 & 0.016184s & 86 & 6 & 0.2112s\tabularnewline
\hline 
HWB (x6) & 7 & 116 & 55 & 227 & $56\pm23$ & 1.479s & 52 & 0.023293s & 126 & 28 & 0.767694s\tabularnewline
\hline 
Hamming (low) & 17 & 236 & 52 & 652 & $121\pm16$ & 4.531s & 44 & 0.056574s & 240 & 41 & 1.447838s\tabularnewline
\hline 
Hamming (medium) & 17 & 534 & 66 & 1247 & $271\pm49$ & 8.593s & 53 & 0.139646s & 554 & 36 & 4.9049s\tabularnewline
\hline 
GF($2^{7}$) mult & 21 & 300 & 196 & 1013 & $145\pm19$ & 6.799s & 149 & 0.057228s & 344 & 109 & 4.441129s\tabularnewline
\hline 
GF(55) mult & 9 & 48 & 30 & 122 & $22\pm6$ & 0.829s & 12 & 0.010317s & 52 & 7 & 0.120845s\tabularnewline
\hline 
MQ-Toffoli (x3) & 5 & 18 & 12 & 34 & $11\pm4$ & 0.238s & 4 & 0.005102s & 20 & 2 & 0.03555s\tabularnewline
\hline 
MQ-Toffoli (x4) & 7 & 30 & 20 & 59 & $15\pm6$ & 0.420s & 8 & 0.007605s & 32 & 2 & 0.078723s\tabularnewline
\hline 
MQ-Toffoli (x5) & 9 & 42 & 28 & 92 & $26\pm9$ & 0.661s & 8 & 0.010962s & 44 & 2 & 0.109068s\tabularnewline
\hline 
MQ-Toffoli (x10) & 19 & 102 & 68 & 254 & $69\pm11$ & 2.001s & 8 & 0.033435s & 104 & 6 & 0.3145s\tabularnewline
\hline 
RC adder (6-qb) & 14 & 93 & 11 & 157 & $31\pm17$ & 1.376s & 11 & 0.01955s & 97 & 6 & 0.209832s\tabularnewline
\hline 
5 mod 4 & 5 & 28 & 19 & 46 & $14\pm5$ & 0.320s & 14 & 0.006838s & 32 & 7 & 0.10531s\tabularnewline
\hline 
C-sum (mux 9) & 30 & 168 & 24 & 424 & $32\pm11$ & 3.267s & 16 & 0.026376s & 174 & 14 & 0.308804s\tabularnewline
\hline 
QFT (4-qubit) & 5 & 46 & 30 & 79 & $25\pm12$ & 0.588s & 20 & 0.013631s & 50 & 8 & 0.212407s\tabularnewline
\hline 
VBE-Adder & 10 & 70 & 20 & 113 & $22\pm7$ & 0.891s & 14 & 0.013919s & 72 & 7 & 0.189217s\tabularnewline
\hline 
Mod-reduce & 11 & 105 & 31 & 175 & $31\pm12$ & 1.403s & 30 & 0.025929s & 111 & 14 & 0.557424s\tabularnewline
\hline 
\end{tabular}
\par\end{centering}
\caption{\label{tab:BenchResultsExtra-1}Extra benchmark results: QISkit and EN compiler.}
\end{sidewaystable*}

\end{document}